\documentclass[twocolumn,showpacs,preprintnumbers,aps,prd,longbibliography,10pt]{revtex4-1}

\usepackage{graphicx,amssymb,amsmath,amsthm,amsfonts,epsfig}
\usepackage[linktocpage]{hyperref}
\usepackage[usenames,dvipsnames]{color}
\usepackage{epstopdf}
\usepackage{cleveref}
\definecolor{coolblack}{rgb}{0.0, 0.18, 0.39}
\definecolor{darkred}{rgb}{0.5,0,0}
\definecolor{darkgreen}{rgb}{0,0.5,0}
\definecolor{darkblue}{rgb}{0,0,0.5}
\definecolor{lapislazuli}{rgb}{0.15, 0.38, 0.61}
\definecolor{venetianred}{rgb}{0.78, 0.03, 0.08}
\definecolor{bleudefrance}{rgb}{0.19, 0.55, 0.91}
\definecolor{dogwoodrose}{rgb}{0.84, 0.09, 0.41}
\definecolor{dogwoodrose}{rgb}{0.84, 0.09, 0.41}
\hypersetup{colorlinks=true, citecolor=darkblue, linkcolor=darkblue, 
urlcolor = darkblue}

\usepackage{amsmath,amssymb}
\usepackage{tensor}

\def\be{\begin{equation}}
\def\ee{\end{equation}}

\newcommand{\bea}{\begin{eqnarray}}
\newcommand{\eea}{\end{eqnarray}}
\newcommand{\ben}{\begin{enumerate}}
\newcommand{\een}{\end{enumerate}}
\newcommand{\bi}{\begin{itemize}}
\newcommand{\ei}{\end{itemize}}

\def\ga{\mathrel{\raise.3ex\hbox{$>$\kern-.75em\lower1ex\hbox{$\sim$}}}}
\def\la{\mathrel{\raise.3ex\hbox{$<$\kern-.75em\lower1ex\hbox{$\sim$}}}}

\def\l{\left}
\def\r{\right}

\def\I_M{{I_{\scriptscriptstyle M\times M}}}

\def\be{\begin{equation}}
\def\ee{\end{equation}}
\def\bea{\begin{eqnarray}}
\def\eea{\end{eqnarray}}
\newcommand{\beq}{\begin{eqnarray}}
\newcommand{\eeq}{\end{eqnarray}}
\def\pa{\partial}

\newcommand{\beqa}{\begin{eqnarray}}
\newcommand{\eeqa}{\end{eqnarray}}

\begin{document}
\title{\large Scattering by regular black holes: Planar massless scalar waves impinging upon a Bardeen black hole}

\author{Caio F. B. Macedo}\email{caiomacedo@ufpa.br}
\author{Ednilton S. de Oliveira}\email{ednilton@pq.cnpq.br}
\author{Lu\'is C. B. Crispino}\email{crispino@ufpa.br}
\affiliation{Faculdade de F\'{\i}sica, Universidade 
Federal do Par\'a, 66075-110, Bel\'em, Par\'a, Brazil.}

\begin{abstract}
Singularities are common features of general relativity black holes. However, within general relativity, one can construct black holes that present no singularities. These regular black hole solutions can be achieved by, for instance, relaxing one of the energy conditions on the stress-energy tensor sourcing the black hole. Some regular black hole solutions were found in the context of nonlinear electrodynamics, the Bardeen black hole being the first one proposed. In this paper, we consider a planar massless scalar wave scattered by a Bardeen black hole. We compare the scattering cross section computed using a partial-wave description with the classical geodesic scattering of a stream of null geodesics, as well as with the semiclassical glory approximation. We obtain that, for some values of the corresponding black hole charge, the scattering cross section of a Bardeen black hole has a similar interference pattern to a Reissner-Nordstr\"om black hole.

\end{abstract}

\maketitle

\section{Introduction}
Black holes (BHs) are among the most interesting objects of general relativity (GR). Although GR is a highly nonlinear theory, BHs come out of it with a very simple structure. Standard GR BH solutions are parameterized by their mass, charge and angular momentum \cite{heuslerreview} (see, e.g., Refs. \cite{Herdeiro:2014goa,Benone:2014ssa} for interesting counterexamples of the previous statement). Although related to the earliest predictions of GR, the strong field regime of BHs is still an experimental challenge \cite{psaltisreview,willreview,Berti:2015itd}. Notwithstanding, the observational data presently available suggest that BHs populate basically all the galaxies in the Universe \cite{Narayan:2005ie}.

Although very successful in explaining the available data, standard GR BHs suffer from one of the main problems of GR: the presence of singularities. Hawking and Penrose indeed showed that, for some hypotheses on the gravitational collapse, the formation of singularities in BHs would be unavoidable \cite{Penrose:1964wq,hawkingb}. These singularities were conjectured by Penrose to be hidden by a horizon \cite{Penrose:1969pc,Wald:1997wa}, and were claimed to be possibly avoided within an improved theory of gravity (extension or modification of GR) \cite{Clifton:2011jh}.

The study of BHs without singularities can help us to understand the role played by singularities in astrophysics. Still within GR, one can obtain BHs without singularities ---  dubbed regular BHs --- by relaxing one of the energy conditions on the stress-energy tensor. Bardeen proposed the first regular BH solution \cite{bardeen}, which was later identified as a solution for a nonlinear magnetic monopole~\cite{AyonBeato:2000zs}. Since then, other regular BHs appeared in the literature, in different scenarios (see, e.g., Refs.~\cite{Ansoldi:2008jw,Lemos:2011dq,Flachi:2012nv}, and references therein). Moreover, regular BHs can be relevant in the context of quantum gravity \cite{Hayward:2005gi,DeLorenzo:2014pta}, and some of them reproduce the quantum weak field regime of GR \cite{BjerrumBohr:2002kt}.

Studies of scattering by BHs have been extensively made \cite{Futterman:1988ni}. Reference \cite{Crispino:2009xt} presented the results  of the scattering of all basic massless (spin 0, 1/2, 1 and 2) fields by Schwarzschild BHs. Moreover, the shadows of BHs \cite{Falcke:1999pj,Huang:2007us} may become visible with future telescopes, like the Event Horizon Telescope \cite{eventhorizon}, and the scattering of light, considering wave and semiclassical approximations, may be important in anticipating subtle characteristics of the shadows. However, a careful study of the scattering of fields by regular BHs is still lacking in the literature~\footnote{An approximation scheme to compute the scattering cross section of regular BHs was performed in Ref.~\cite{Huang2014}. The scheme relies on the Wentzel-Kramers-Brillouin (WKB) approximation, using a modification of the effective potential \eqref{eq:pot}, leading to inaccurate results (see, e.g., Ref. \cite{Macedo:2014uga}).}.

The line element of the Bardeen BH can be written as	
\be
ds^2=-f(r)dt^2+f(r)^{-1}dr^2+r^2(d\theta^2+\sin^2\theta d\varphi^2),
\label{eq:ds}
\ee
where the lapse function $f(r)$ is given by
\be
f(r)=1-\frac{2Mr^2}{(r^2+q^2)^{3/2}}.
\ee
The Bardeen BH has a structure similar to the Reissner-Nordstr\"{o}m (RN) BH (see, e.g., Ref. \cite{Ansoldi:2008jw}). For $q<q_{\rm ext}=4M/(3\sqrt{3})$ (henceforth, without loss of generality, we shall assume $q\geq 0$) the spacetime has two horizons and for $q=q_{\rm ext}$ the horizons degenerate, characterizing the extremal case. Following Ref. \cite{Macedo:2014uga}, we shall display our results in terms of the normalized charge $Q=q/{q_{\rm ext}}$~\footnote{The parameter presented here as $Q$ can also be interpreted as a natural length when one considers that the Bardeen BH comes from a quantum generalization of GR. See, for instance, Ref. \cite{DeLorenzo:2014pta} for more details.}.

\begin{figure}%
\includegraphics[width=\columnwidth]{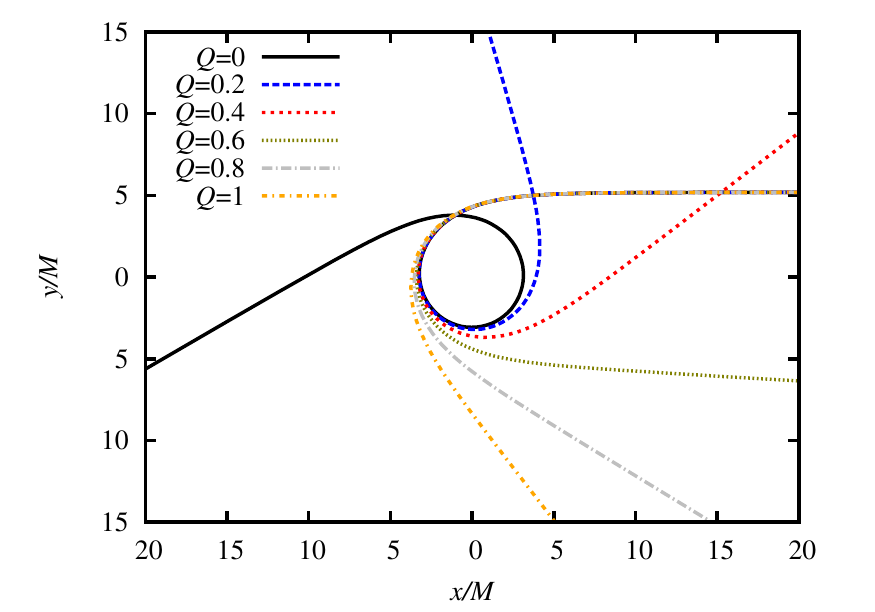}%
\caption{Geodesics approaching a Bardeen BH from infinity with an impact parameter of $b=5.2M$, for different values of the BH charge. The Schwarzschild case ($Q=0$) is also exhibited (solid line).}%
\label{fig:geo_cha}%
\end{figure}

It is interesting to note that the variation of the charge of the regular BH changes considerably the deflection angle of light rays \cite{Zhou:2011aa}. In Fig. \ref{fig:geo_cha} we plot null geodesics coming from infinity with an impact parameter of $b=5.2M$, for different values of the BH charge. The behavior is qualitatively similar to the case of RN BHs \cite{Crispino_2009:prd79_064022}. We see that, adjusting the BH charge, we can have scattering in basically any direction.

The remainder of this paper is organized as follows: In Sec.~\ref{sec:classical} we review the main aspects of the classical geodesic scattering and semiclassical glory approximation to compute the differential scattering cross section. In Sec.~\ref{sec:planar} we present the partial-wave method used to compute the scattering cross section of planar massless scalar waves. In Sec.~\ref{sec:results} we present the results for the scattering of planar massless scalar waves impinging  upon a Bardeen regular BH, comparing the three different approaches used to compute the scattering cross section.  In Sec.~\ref{sec:final} we end up with our final remarks. Throughout this work we use $G=c=\hbar=1$ and metric signature $(-,+,+,+)$.

\section{Classical scattering and semiclassical glory} \label{sec:classical}
In this section we investigate the scattering by BHs by using two approaches: classical geodesic scattering and the semiclassical glory approximation. These approaches allow us to foresee some of the aspects of the scattering cross section obtained within the full partial-wave analysis.
\subsection{Geodesic scattering}

The analysis of null geodesics in the Bardeen spacetime can be seen in Ref. \cite{Macedo:2014uga}. For the classical approximation of the scattering we may consider a stream of parallel null geodesics coming from infinity. In this case, the analysis of Ref. \cite{Collins:1973xf} suits the problem of classical scattering by Bardeen BHs. The classical scattering cross section is given by
\be
\frac{d\sigma}{d\Omega}=\frac{1}{\sin\chi}\sum{b(\chi)\l|\frac{db(\chi)}{d\chi}\r|},
\label{eq:classicalsca}
\ee
where $b(\chi)$ is the impact parameter associated with a scattering angle $\chi$. The summation in Eq. \eqref{eq:classicalsca} is such that we also consider the case in which the null geodesic rotates (one or many times) around the BH before going to infinity (for instance, see the solid curve of Fig. \ref{fig:geo_cha}). It is interesting to note that the classical scattering formula given by Eq. \eqref{eq:classicalsca} describes very well the planar-wave case for small scattering angles, although it gives discrepant results for moderate-to-high scattering angles, as we shall see in Sec. \ref{sec:results}.

Let us now obtain $b(\chi)$ through a geodesic analysis.
Without loss of generality, we shall restrict the geodesic motion to the plane $\theta = \pi/2$. From the line element~\eqref{eq:ds}, we can write, for null
geodesics
\be
\left( \frac{du}{d\varphi} \right)^2 = \frac{1}{b^2} - f(1/u) u^2,
\label{eq:orbit}
\ee
where we have defined $u \equiv 1/r$, $b \equiv L/E$ is the impact
parameter given in terms of the constants of motion
\be
E = f \dot{t} \qquad \text{and} \qquad L = r^2 \dot{\varphi},
\label{eq:constants}
\ee
and the overdot denotes differentiation with respect to an affine
parameter of the curve.

Differentiating Eq.~\eqref{eq:orbit} with respect to $\varphi$, we obtain
\be
\frac{d^2 u}{d\varphi^2} = - \frac{u^2}{2} \frac{df(1/u)}{du} - uf(1/u).
\label{eq:orbit2}
\ee
Solving Eq.~\eqref{eq:orbit2} with the appropriate boundary conditions,
one can obtain the geodesics followed by massless particles, such as the ones shown in
Fig.~\ref{fig:geo_cha}. 
The smallest positive root of the right-hand side of Eq.~\eqref{eq:orbit2}
corresponds to the radius of the critical
orbit for null geodesics, $u_c = 1/r_c$. Substituting its value in the
right-hand side of Eq.~\eqref{eq:orbit} and setting it to zero,
we obtain the impact parameter associated with the critical orbits,
$b_c$. Going in the other way around, i.e., choosing a value $b > b_c$,
the smallest root of the right-hand side of Eq.~\eqref{eq:orbit}
is $u_0 = 1/r_0$, where $r_0$ is the turning point --- or the radius of
maximum approximation --- of the geodesic.

\begin{figure*}
\includegraphics[width=\columnwidth]{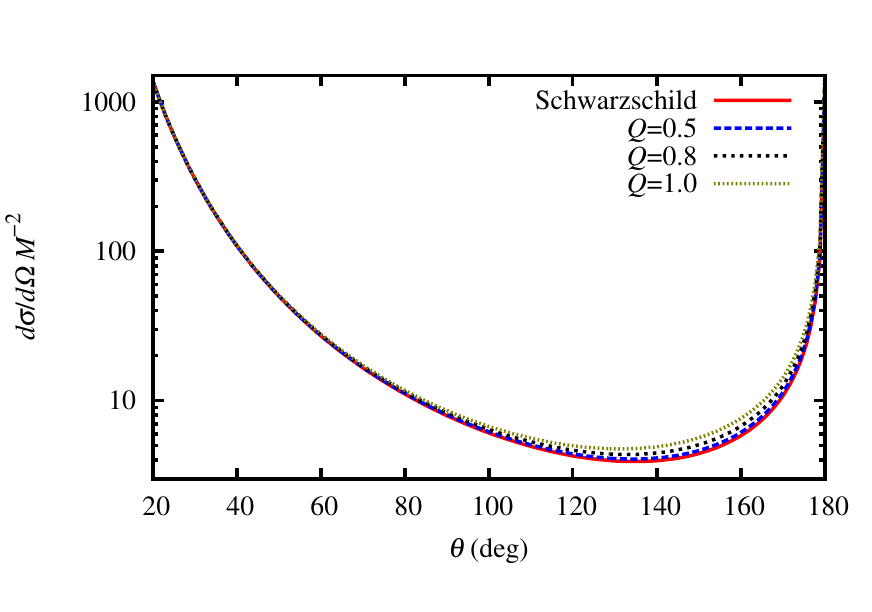}
\includegraphics[width=\columnwidth]{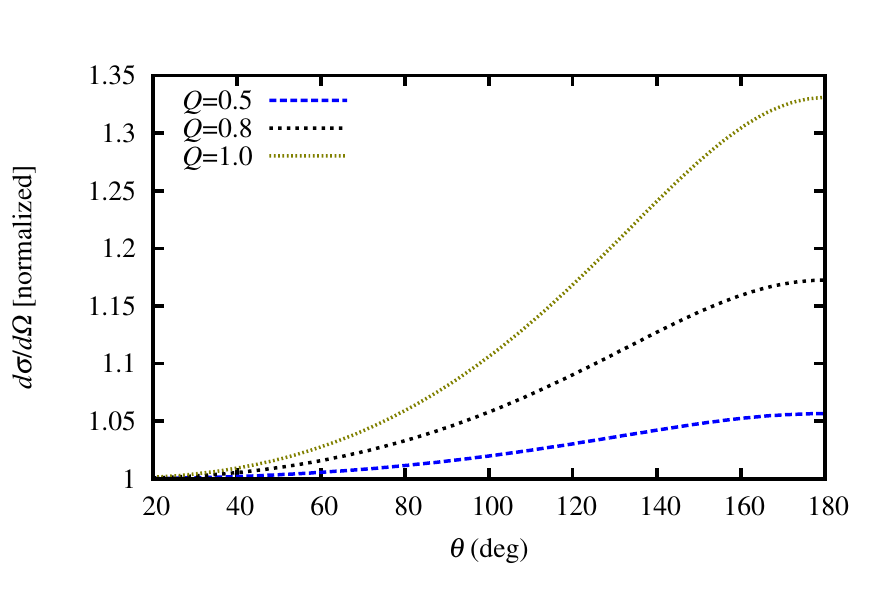}
\caption{\textit{Left panel:} Classical scattering cross section of  Bardeen BHs, with $Q=0.5,\,0.8$ and $1$, and for the Schwarzschild BH ($Q=0$). \textit{Right panel:} Classical scattering cross section of Bardeen BHs, normalized by the Schwarzschild case.}%
\label{fig:classgeo}%
\end{figure*}

Finally, by integrating Eq.~\eqref{eq:orbit} in the case of
scattered geodesics, we obtain
\be
\alpha = \int\limits_{0}^{u_0} \left[\frac{1}{b^2} - f(1/u) u^2\right]^{-1/2} du.
\label{eq:alpha}
\ee
The deflection angle following directly from Eq.~\eqref{eq:alpha} is given by
\be
\Theta(b) = 2\alpha(b) - \pi.
\label{eq:Theta}
\ee
By inverting Eq.~\eqref{eq:Theta}, one obtains $b(\Theta)$, and substituting it in Eq. \eqref{eq:classicalsca}, one obtains the classical scattering cross section. Plots of the classical scattering cross section obtained by Eq. \eqref{eq:classicalsca} are exhibited in Fig. \ref{fig:classgeo}, where we compare the Bardeen cases with the Schwarzschild one. We see from Fig. \ref{fig:classgeo} that the increase of the BH charge contributes to an increase of the BH classical scattering cross section.

\subsection{Glory scattering}\label{sec:glory}
The interference that occurs between scattered partial waves with different angular momenta is not taken into account by the classical formula~\eqref{eq:classicalsca}.
In order to obtain a scattering cross section that takes into account
the interference processes, we need to perform a wave analysis of the problem.
Before going into the full wave analysis, however, it is interesting to apply an approximate method that works remarkably well for high
scattering angles ($\theta \sim \pi$) and that captures
some key features of the scattering cross section in this regime, including the
interference process: the semiclassical glory approximation
\cite{PhysRevD.31.1869}. Indeed, one of the main advantages of this semiclassical approximation is that one can
find an analytical formula that gives some physical insight for the
width of interference fringes in the scattering cross section as well
as the intensity of the scattered flux for $\theta\sim \pi$.
As a semiclassical approximation, it is valid for
$\omega M \gg 1$, although it can still reproduce remarkably well some results for~$\omega M \sim 1$.

The semiclassical formula for the glory scattering by spherically symmetric BHs is given by \cite{PhysRevD.31.1869}
\be
\frac{d\sigma_{\rm sc}}{d\Omega}=2\pi \omega b_g^2 \l|\frac{db}{d\theta}\r|_{\theta = \pi} J_{2s}^2(\omega b_g\sin\theta),
\label{eq:glory}
\ee
where $b_g$ is the impact parameter of backscattered rays
($\theta =\pi$), $J_{2s}(x)$ is the Bessel function of the first kind
(of order $2s$), and $s$ is the wave spin. In our case, since we are considering a scalar wave, $s=0$. We note that
there are multiple values of $b_g$ corresponding to the multiple values
of the deflection angle, namely $\Theta = \pi + 2n\pi$, with $n = 0, 1, 2 \dots$,
that result on back-scattered rays.
All the rays scattered close to the the backward direction ($\theta\sim\pi$) contribute to the
glory scattering, but the most important contribution comes from the
$n = 0$ case. The next
contribution, $ n = 1$, has an intensity that is about 0.2 \%
of the $n=0$ one in the Schwarzschild case, and
about 0.8 \% in the case of the extreme Bardeen BH.
This is a consequence of the derivative $|db/d\theta|_{\theta = \pi}$ in Eq. \eqref{eq:glory} getting rapidly suppressed as $n$ increases. In fact,
the values of $b_g$ for rays that pass multiple times around the
BH are very close to each other and also to $b_c$. Here, we 
consider only the most important contribution to the glory
scattering.

\begin{figure}
 \centering
 \includegraphics[width=\columnwidth]{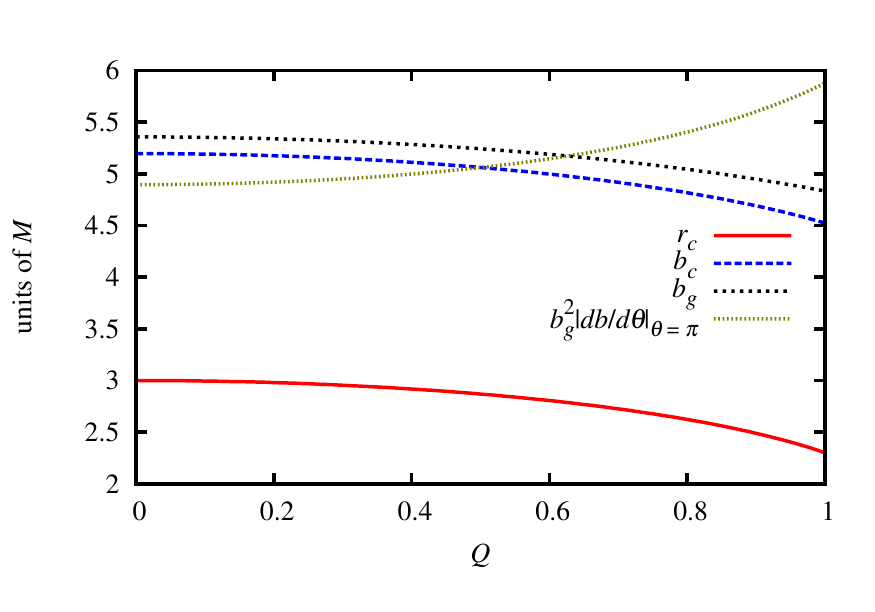}
 \caption{Glory scattering parameters for Bardeen BHs
 with varying charge, considering only the dominant contribution ($\Theta = \pi$).
 We note that, with the exception of $|db/d\theta|_{\theta = \pi}$, all
 the important parameters related to the glory scattering decrease monotonically with
 the increase of $Q$.}
 \label{fig:glory_params}
\end{figure}

Once we have the knowledge of the glory scattering formula, Eq.~\eqref{eq:glory}, we only
need to determine $b_g$ and $|db/d\theta|_{\theta = \pi}$ in order to
obtain the glory scattering cross section. 
Therefore,
we apply Newton's method and numerical integration to obtain the
parameters $b_g$ and $|db/d\theta|_{\theta = \pi}$. Numerical results
for $r_c$, $b_c$, $b_g$ and $b_g^2|db/d\theta|_{\theta = \pi}$ are
presented in Fig.~\ref{fig:glory_params}. From these results and
Eq.~\eqref{eq:glory}, we may expect that (i) interference
fringes get wider and (ii) backscattered flux intensity is
enhanced for higher values of the BH charge. Expectation (i) comes from
the fact that the interference fringe width is inversely proportional
to $b_g$, as indicated by the argument of the Bessel function in Eq.~\eqref{eq:glory}. Moreover, Fig.~\ref{fig:glory_params} shows that
$b_g$ decreases monotonically as $Q$ increases. We also note from
Eq.~\eqref{eq:glory} that the scattering intensity is proportional
to $b_g^2|db/d\theta|_{\theta = \pi}$, and to the wave frequency. 
As shown in Fig.~\ref{fig:glory_params},
although $b_g$ decreases monotonically with the increase of $Q$,
$b_g^2|db/d\theta|_{\theta = \pi}$ increases monotonically with $Q$, 
justifying expectation (ii).

The above analysis may be compared with the results for the glory
scattering from RN BHs~\cite{Crispino_2009:prd79_064022}.
In the latter case, $b_g$ decreases monotonically with the increase of
the BH charge, while $|db/d\theta|_{\theta = \pi}$ increases monotonically (\emph{cf.} Fig. 9 of Ref.~\cite{Crispino_2009:prd79_064022}).
Therefore, considering the change of the BH charge,
the behavior of the parameters $b_g$ and $|db/d\theta|_{\theta = \pi}$
are qualitatively the same for Bardeen and RN BHs.
In the case of RN BHs, however, the combination
$b_g^2|db/d\theta|_{\theta = \pi}$ does not
increase monotonically with the charge --- as it happens
for Bardeen BHs. Instead, the glory scattering amplitude as a function of $Q$ presents a local minimum in the RN case (cf. Fig. 10 of Ref.~\cite{Crispino_2009:prd79_064022}).

In Sec.~\ref{sec:results}, we compare results obtained from
Eq.~\eqref{eq:glory} with partial-wave results, exhibiting excellent
agreement in the regime $\theta \lesssim \pi$.

\section{Planar wave scattering} \label{sec:planar}

Planar massless scalar waves, represented by the wave function $\Phi$, are described by the Klein-Gordon equation
\be
 \frac{1}{\sqrt{-g}}\pa_a \l(\sqrt{-g}g^{ab}\pa_b\Phi\r)=0.
\label{eq:kleingordon}
\ee
Here we shall be interested in monochromatic plane waves. We have
\be
\Phi_{\omega} = \sum_{lm} \frac{\phi(r)}{r} Y^m_l(\theta,\varphi) e^{-i \omega t},
\label{eq:expansion}
\ee
where $Y^m_l(\theta,\varphi)$ are the scalar spherical harmonics. Substituting Eq. \eqref{eq:expansion} into Eq. \eqref{eq:kleingordon}, we obtain the following radial equation:
\be
\l(-\frac{d}{dr_*^2}+V_\phi(r) -\omega^2\r)\phi(r)=0,\label{eq:eqr}
\ee
in which $r_*$ is the tortoise coordinate, defined through $dr_*= f(r)^{-1} dr$, and 
\be
V_\phi(r)=f\left(\frac{l (l+1)}{r^2}+\frac{f'}{r}\right)\label{eq:pot}
\ee
is the scalar field effective potential. The scalar field potential is localized, going to zero at both asymptotic limits of $r_*$ (infinity and horizon) \cite{Macedo:2014uga}.

Plane waves coming from infinity can be described in terms of the so-called $in$ modes. These modes are purely incoming from the past null infinity, obeying the following boundary conditions:
\be
\phi(r)\sim\l\{
\begin{array}{ll}
R_I+\mathcal{R}_{\omega l}R_I^{*},&{\rm as}~ r_*\to +\infty ~(r\to +\infty),\\
  \mathcal{T}_{\omega l} R_{II}, &{\rm as}~r_*\to - \infty ~(r\to r_h),
\end{array}\r.
\label{eq:inmodes}
\ee
with
\bea
R_I&=&e^{-i \omega r_*}\sum_{j=0}^N \frac{A^{(j)}_\infty}{r^j},\label{eq:ri}\\
R_{II}&=&e^{-i \omega r_*}\sum_{j=0}^N (r-r_h)^j A^{(j)}_{r_h},\label{eq:rii}
\eea
where $|\mathcal{R}_{\omega l}|^2$ and $|\mathcal{T}_{\omega l}|^2$ are the reflection
and transmission coefficients, respectively.
Flux conservation implies that
$|\mathcal{R}_{\omega l}|^2 + |\mathcal{T}_{\omega l}|^2 = 1$.
Note that the summations in Eqs. \eqref{eq:ri} and \eqref{eq:rii} are required to keep track of the convergence of the solutions. The numerical infinity and horizon are chosen such that $V_\phi(r)\ll \omega^2$ at the boundaries.

The scalar differential scattering cross section for Bardeen BHs can be
written in terms of partial waves as~\cite{Futterman:1988ni}
\be
\frac{d\sigma}{d\Omega} = |g(\theta)|^2,
\label{eq:sig}
\ee
where
\be
g(\theta) = \frac{1}{2 i \omega} \sum\limits_{l = 0}^{\infty}(2l+1)\left[e^{2i\delta_l(\omega)} - 1\right] P_l(\cos\theta)
\label{eq:amp}
\ee
is the scattering amplitude, with the phase shifts ($\delta_l$) given
by
\be
e^{2i\delta_l(\omega)} \equiv (-1)^{l+1} \mathcal{R}_{\omega l}.
\label{eq:def}
\ee

\section{Results}\label{sec:results}

In order to obtain the phase shifts to compute the scattering cross section
via the partial-wave method, we have applied a fourth-fifth Runge-Kutta
method to solve the radial equation \eqref{eq:eqr}. We have typically
started with the near-horizon condition at $r_s = 1.0001 r_h$, and the outer boundary (numerical infinity) chosen depends on the value of $l$. Results were
obtained with boundary conditions~\eqref{eq:inmodes}, as well as with
alternative conditions in terms of spherical Hankel functions
(see, e.g., Eq.~(18) of Ref.~\cite{Crispino_2009:prd79_064022}).
Both conditions lead basically to the same results. Since the
sum in Eq.~\eqref{eq:amp} does not converge very quickly, because of the Coulomb
characteristic of the problem, we have applied the convergence method first
introduced by Yennie \emph{et al.}~\cite{Yennie_1954:prd95_500}, and first applied to the BH scattering problem by Dolan \textit{et al.} in Ref. \cite{Dolan:2006vj}.

\begin{figure}%
\includegraphics[width=\columnwidth]{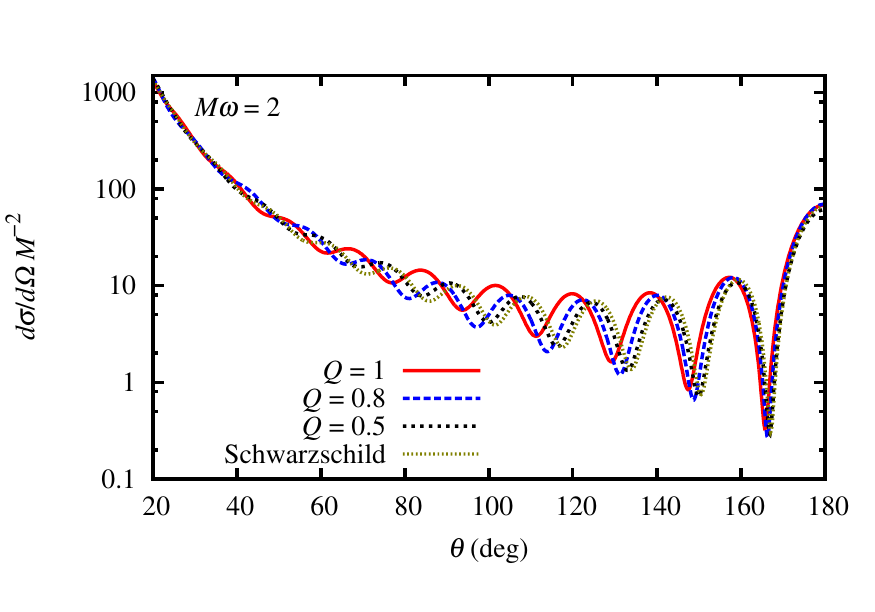}%
\caption{Scattering cross sections for Bardeen BHs considering different BH charges. We also plot the Schwarzschild case, for comparison. We see that the value of the charge affects the fringe widths, while the (avarage) amount of scattered flux remains basically the same.}%
\label{fig:scat_cha}%
\end{figure}

\begin{figure*}
\includegraphics[width=\columnwidth]{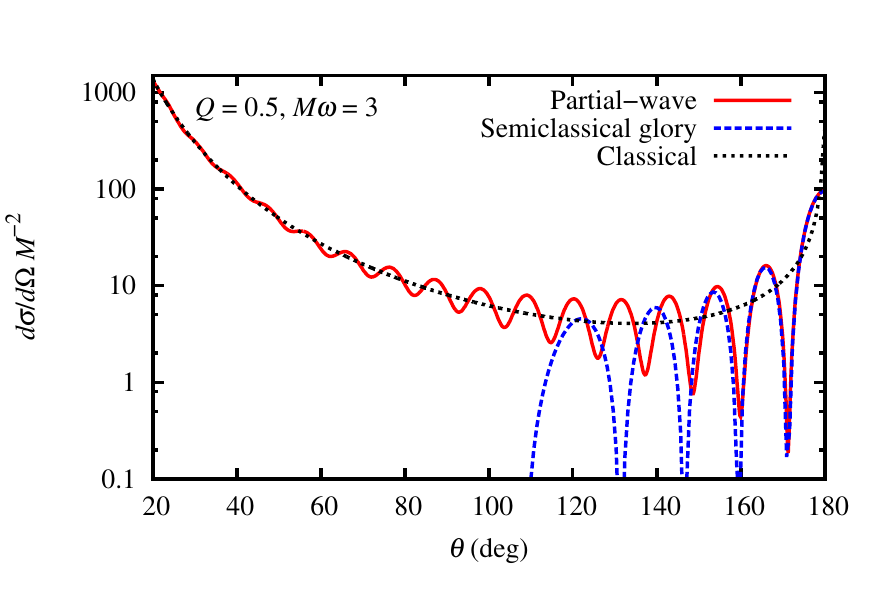}
\includegraphics[width=\columnwidth]{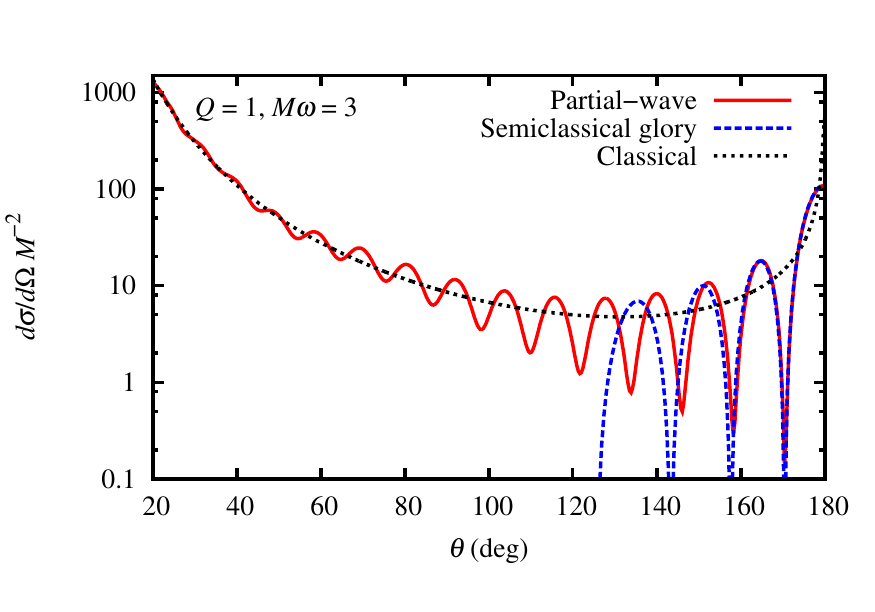}
\caption{Comparison of partial-wave, semiclassical glory and classical geodesic approaches for the differential scattering cross section, for $M\omega=3$ (in the first two cases), and different values of the Bardeen BH charge [$Q=0.5$ (left) and $Q=1$ (right)]. The semiclassical glory approximation reproduces very well the results for backscattered waves ($\theta\sim \pi$), while the classical approach works well for small scattering angles.}%
\label{fig:compari}%
\end{figure*}

In Fig. \ref{fig:scat_cha} we show the scattering cross section for
Bardeen BHs with different charges ($Q = 0.5, 0.8, 1$), as well as for the Schwarzschild BH, and
$M \omega = 2$.
We see that the fringe widths increase with the increase of the BH charge.
This, as anticipated by the semiclassical analysis of Sec. \ref{sec:glory}, is in accordance with the fact that $b_g$ decreases monotonically
with the increase of $Q$, as previously seen in Fig.~\ref{fig:glory_params}.
The amount of scattered flux (on average) remains basically the same. These general behaviors are similar to the ones presented by the RN BHs \cite{Crispino_2009:prd79_064022}.

\begin{figure}%
\includegraphics[width=\columnwidth]{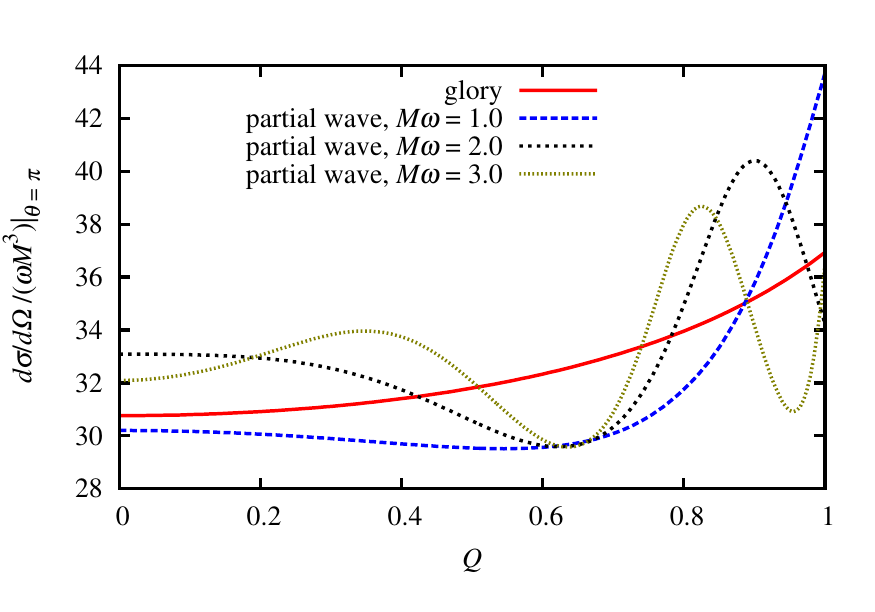}%
\caption{Glory intensity in the backward direction (normalized by the wave frequency) ($\theta=\pi$) as a function of the Bardeen BH charge. We see that the amplitude computed through the partial-wave method oscillates around the one computed through the glory approximation.}%
\label{fig:scat_back}%
\end{figure}

Figure~\ref{fig:compari} presents comparisons of the numerical
scalar scattering cross sections for Bardeen BHs with the
approximated geodesic and glory results. We see that the glory
results fit remarkably well the numerical results for large angles
($\theta \lesssim \pi$), while the geodesic results fit well the small-angle region. This very good agreement can also
be considered as a consistency check of our results.

The glory approximation can be used to capture most of the features of the back-scattered wave. Some caution, however, should be taken when one considers the glory intensity in the backward direction (normalized by the wave frequency). To illustrate this, in Fig. \ref{fig:scat_back} we plot the amplitudes of the back-scattered wave, for $M\omega=1,\,2$ and $3$, computed through the partial-wave method and through the glory approximation, as a function of the Bardeen BH charge. We see that the results obtained via the partial-wave method oscillate around the one obtained using the glory approximation. This agrees with the analysis presented in Ref. \cite{Crispino_2009:prd79_064022} for RN BHs.

\begin{figure*}
\includegraphics[width=\columnwidth]{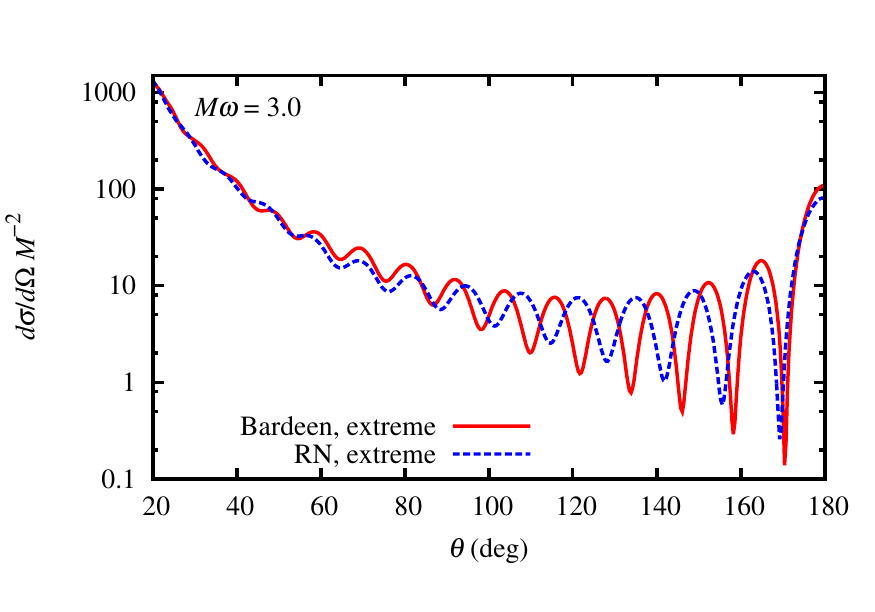}
\includegraphics[width=\columnwidth]{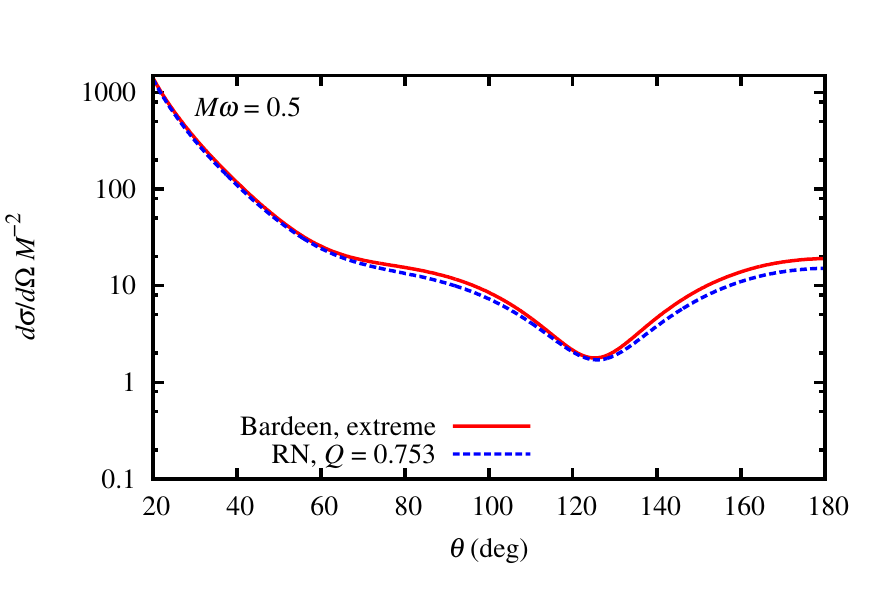}\\
\includegraphics[width=\columnwidth]{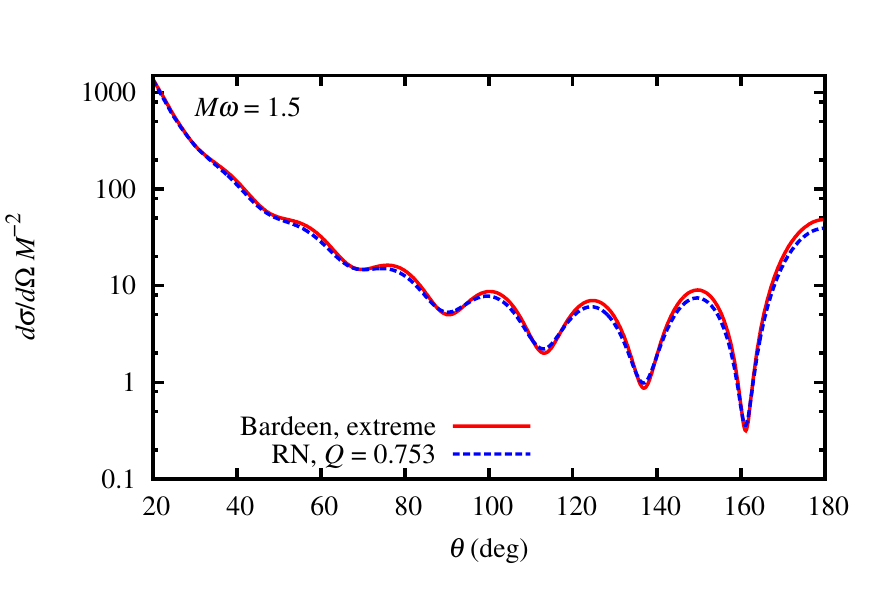}
\includegraphics[width=\columnwidth]{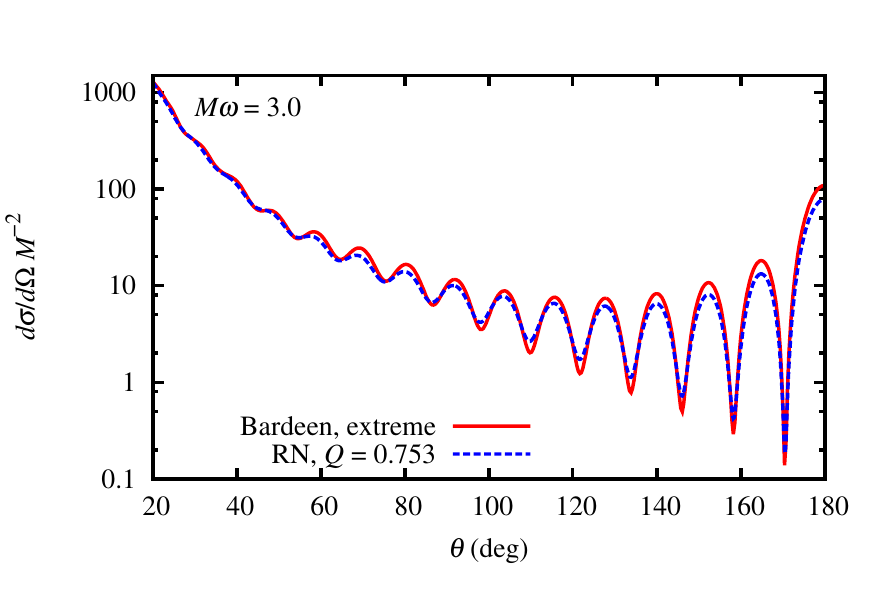}
\caption{Comparison between Bardeen and RN BH scalar scattering.  
\textit{Top-left panel}: The case of extremal Bardeen and RN BHs
for $M\omega = 3.0$.
\textit{Top-right panel}: Bardeen extreme BH
scattering compared with the scattering from a RN BH with
$Q = 0.753$, for $M\omega = 0.5$.
\textit{Bottom panels}: The same as the top-right panel, but with $M\omega  = 1.5$ (left) and $M\omega=3.0$ (right).
}
\label{fig:bd_rn-comp}
\end{figure*}

In Fig.~\ref{fig:bd_rn-comp} we compare the differential scattering cross sections of Bardeen and RN BHs. While RN and Bardeen BHs with the correspondent charge produce different scattering patterns --- illustrated by the top-left panel of Fig. \ref{fig:bd_rn-comp} --- we can have configurations with different charges that produce almost the same scattering pattern. A similarity of the patterns also happens in the case of absorption cross sections, when the critical impact parameter ($b_c$) of the RN and Bardeen cases are the same 
\cite{Macedo:2014uga}. Here, however, the similarity of the scattering cross sections intensifies when $b_g$ for the RN and Bardeen cases match. The similarities are illustrated in the top-right and bottom panels of Fig. \ref{fig:bd_rn-comp}, where we show the scattering cross section for a RN BH with $Q=0.753$ and for a Bardeen BH with $Q=1$, for different values of the frequency. The scattering flux intensities
are different for intermediate-to-high scattering angles, while the interference widths are essentially the same for all scattering angles.

\section{Final remarks} \label{sec:final}

We have computed the scalar scattering cross section of
regular Bardeen BHs. Numerical results were compared with
both geodesic and glory approximations, and we have found excellent
agreement within the validity limits of each approximation.

From the glory approximation, it is known that the interference
fringe widths depend inversely on the impact parameter of
backscattered waves, $b_g$. The classical analysis from geodesics shows
that $b_g$ decreases monotonically with the increase of $Q$.
Therefore, we expect that the interference fringes get wider as
$Q$ increases. This was confirmed from our numerical results
obtained via partial-wave method. 

Comparison of Bardeen BHs with RN BHs reveals that the
scattering of these two kinds of BHs can be similar but not
identical. By similar we mean that in both cases the scattering
cross section presents (i) intense oscillations in the near-backward
scattering, (ii) rapidly growing flux amount and smoother oscillations for
smaller angles, and (iii) similar results for very small scattering 
angles. (i) is a consequence of the strong interference between
rays passing by the opposite sides of the BH, as it is well
described by the glory approximation in both
cases.
(ii) is a consequence of the fact that for small scattering angles both $b$ and
$|db/d\theta|$ increase as $\theta$ diminishes, and the
difference between paths followed by neighboring rays becomes smaller,
weakening interference effects. We can conclude that the main
contribution to the scattering cross section for very small angles
comes from rays with high impact parameters~\footnote{There are also
contribution from rays with $b \sim b_c$ that pass multiple times
around the BH, but since $|db/d\theta|$ is very small in these
cases, their contribution to the scattering cross section can be neglected.}. We may treat such cases in the weak-field regime, where
the main contribution to the gravitational interaction comes from
the BH mass, i.e., $f(r) \sim 1 - 2M/r + \mathcal{O}(r^{-n})$,
where $n = 2$ in the case of RN BHs, and $n = 3$
in the case of Bardeen BHs. This explains (iii), i.e., why, for BHs with the same mass, in the regime of small angles, all results tend to be the same, regardless of the nature and value of their charge.

The results presented in this paper reinforce in a way the results presented in Ref.~\cite{Macedo:2014uga}, implying the conclusion that some properties of Bardeen BHs can be very similar to those of RN BHs (with different charge). In this sense, we conclude that it may be difficult to discriminate regular BHs from the standard ones, as far as absorption and scattering of scalar plane waves are concerned.
It should be interesting to extend the analyses presented here
and in Ref.~\cite{Macedo:2014uga} to the scattering and absorption of waves with
higher spins and compare with recently obtained results for RN
BHs~\cite{co2008,cho2009,chm2010,Oliveira:2011zz,cdho2014}.

\begin{acknowledgments}
C.M. thanks A. Flachi for discussions and for pointing out useful references. The authors would like to thank Conselho Nacional de Desenvolvimento Cient\'ifico e Tecnol\'ogico (CNPq), Coordena\c{c}\~ao de Aperfei\c{c}oamento de Pessoal de N\'ivel Superior (CAPES), Funda\c{c}\~ao Amaz\^onia de Amparo a Estudos e Pesquisas do Par\'a (FAPESPA) (FAPESPA), and  Marie Curie action NRHEP-295189-FP7-PEOPLE-2011-IRSES for partial financial support.
\end{acknowledgments}
%
\bibliography{refs_regular}
\end{document}